# Topology-Dependent Brownian Gyromotion of a Single Skyrmion

Le Zhao[1,2,*], Zidong Wang[1,2,*], Xichao Zhang[3,*], Xue Liang[3,*], Jing Xia[3], Keyu Wu[1,2], Heng-An Zhou[1,2], Yiqing Dong[1,2], Guoqiang Yu[4], Kang L. Wang[5], Xiaoxi Liu[6], Yan Zhou[3,†] and Wanjun Jiang[1,2,†]

[1]*State Key Laboratory of Low-Dimensional Quantum Physics and Department of Physics, Tsinghua University, Beijing 100084, China*

[2]*Frontier Science Center for Quantum Information, Tsinghua University, Beijing 100084, China*

[3]*School of Science and Engineering, The Chinese University of Hong Kong, Shenzhen, Guangdong 518172, China*

[4]*Beijing National Laboratory for Condensed Matter Physics, Institute of Physics, Chinese Academy of Sciences, Beijing 100190, China*

[5]*Department of Electrical Engineering, University of California, Los Angeles, California, 90095, USA*

[6]*Department of Electrical and Computer Engineering, Shinshu University, 4-17-1 Wakasato, Nagano 380-8553, Japan*

[*] These authors contributed equally to this work.
[†] To whom correspondence should be addressed:
zhouyan@cuhk.edu.cn and jiang_lab@tsinghua.edu.cn

(Dated: June 2nd, 2020)

**Abstract**

Non-interacting particles exhibiting Brownian motion have been observed in many occasions of sciences, such as molecules suspended in liquids, optically trapped microbeads, and spin textures in magnetic materials. In particular, a detailed examination of Brownian motion of spin textures is important for designing thermally stable spintronic devices which motivates the present study. In this Letter, through using temporally and spatially resolved polar magneto-optic Kerr effect (MOKE) microscopy, we have experimentally observed the thermal fluctuation-induced random walk of a single isolated Néel-type magnetic skyrmion in an interfacially asymmetric Ta/CoFeB/TaO$_x$ multilayer. An intriguing topology dependent Brownian gyromotion behavior of skyrmions has been identified. The onset of Brownian gyromotion of a single skyrmion induced by the thermal effects, including a nonlinear temperature-dependent diffusion coefficient and topology-dependent gyromotion are further formulated based on the stochastic Thiele equation. The experimental and numerical demonstration of topology-dependent Brownian gyromotion of skyrmions can be useful for understanding the nonequilibrium magnetization dynamics and implementing spintronic devices.

PACS: 75.60.Ch, 75.70.Kw, 75.78.-n, 12.39.Dc

Keywords: skyrmion, Brownian motion, topology, spintronics, micromagnetics

Skyrmions are particle-like topological spin textures stabilized by the Dzyaloshinskii-Moriya interaction (DMI)(*1-10*), which have recently stimulated great interests in spintronics community(*11-18*). Specifically, the topological properties of skyrmions are governed by the skyrmion number $Q = 1/4\pi \int \bm{m} \cdot (\partial \bm{m}/\partial x \times \partial \bm{m}/\partial y) \, dxdy$ with $\bm{m} = \bm{M}/M_\mathrm{S}$ being the reduced magnetization vector and $M_\mathrm{S}$ the saturation magnetization, respectively. This quantity is responsible for many intriguing topological transport phenomena(*19-23*). However, an experimental demonstration of topology-dependent Brownian dynamics of skyrmions(*24-29*), is not addressed. This important aspect could facilitate an accurate understanding of the nonequilibrium thermodynamics of skyrmions(*7, 13, 30*). Further, a comprehension of Brownian dynamics could set up a vital step towards thermally stable skyrmion devices(*31-33*), and implementation of skyrmions for novel applications(*34, 35*).

**Simulations of Brownian gyromotion of a single skyrmion.** Brownian motion is a nonequilibrium thermodynamic phenomenon that is typified by the random walk of objects, as a result of thermal fluctuations(*36-38*). In magnetic materials, spin dynamics can be affected by thermal fluctuations, especially when the magnetic anisotropy energy is comparable with the thermal excitations(*39-42*). By incorporating an irregular Gaussian stochastic fluctuating field $\bm{h}(\bm{x}, t)$, one could use the stochastic Landau-Lifshitz-Gilbert (LLG) equation to numerically study the random walk of spin textures, as discussed in Supplemental Material(*43*). The micromagnetic simulation results of isolated Néel-type skyrmions with opposite topological charges ($Q = \pm 1$) are shown in Fig. 1. In our case, the $Q = -1$ skyrmion is defined as the magnetization in center of skyrmion is negative ($m_z < 0$), and $Q = +1$ skyrmion corresponds to the positive orientation in the center ($m_z > 0$), which can be achieved by inverting perpendicular magnetic fields ($\pm B_z$) since $Q$ is an odd function of $\bm{m}$. This feature has been previously used to reveal the skyrmion Hall effect(*22*).

From the time-dependent trajectory, the random-walk feature of skyrmion driven by thermal fluctuation is revealed. More interestingly, we have identified a topology-dependent gyromotion of skyrmions from being clockwise ($Q = +1$) to counterclockwise ($Q = -1$), as shown in Figs. 1(A) and 1(B), respectively. The amplitude of trajectories increases following the increase of temperature, suggesting an enhanced thermally induced diffusion that is consistent with the prediction of conventional Brownian motion. In our simulation, the size of the skyrmion increases following the increase of temperature, which results in an enhanced skyrmion-edge interaction and dissipation.

In a non-confined two-dimensional (2D) geometry, the probability distribution $P(x, y, t)$ of a free Brownian particle at position $(x, y)$ and at time $t$ is described by a Gaussian function $P(x, y, t) = 1/\sqrt{4\pi \mathfrak{D}_{dc} t} \cdot \exp[-(x^2 + y^2)/4\mathfrak{D}_{dc} t]$, where $\mathfrak{D}_{dc}$ is the diffusion coefficient. The value of $\mathfrak{D}_{dc}$ can be estimated by calculating the mean-squared displacement (MSD) with the relation among the MSD, $\mathfrak{D}_{dc}$, and $t^*$ being given as(*36, 38*):

$$\text{MSD}(t^*) = \langle [\boldsymbol{r}_{x,y}(t+t^*) - \boldsymbol{r}_{x,y}(t)]^2 \rangle = 4\mathfrak{D}_{dc} t^* \qquad (1)$$

where $\boldsymbol{r}_{x,y}$ is the position of the skyrmion center in Cartesian coordinates $(x, y)$, $t^*$ is the time interval between the two selected data points. Eq. (1) suggests that the MSD of a Brownian particle should be linearly proportional to $t^*$. However, the extracted MSD from our simulation also exhibits an oscillatory feature that corresponds to the gyrodiffusion of skyrmions, as shown in Fig. 1(C). On the other hand, the diffusion coefficient $\mathfrak{D}_{dc}$ of skyrmion in the linear regime reads as(*28, 35*):

$$\mathfrak{D}_{dc} = k_\text{B} T \frac{\alpha \mathcal{D}}{\mathcal{G}^2 + (\alpha \mathcal{D})^2} \qquad (2)$$

Eq. (2) thus predicts a linear dependence of $\mathfrak{D}_{dc}$ *vs. T* for the fixed values of $\alpha$, $\mathcal{G}$ and $\mathcal{D}$ (*i.e.*, a rigid skyrmion profile with temperature-independent skyrmion diameter $r_\text{sk}$ and domain wall width $\gamma_\text{dw}$ values). Here $k_\text{B}$ is Boltzmann constant, $\alpha$ is the damping coefficient, $\mathcal{G} = 4\pi Q \mu_0 M_\text{S} d/\gamma_0$ is the gyrocoupling vector and $\mathcal{D} = (\mu_0 M_\text{S} d/\gamma_0) \cdot (\pi^3 r_\text{sk}/\gamma_\text{dw})$ is the dissipative tensor with $d$, $\gamma_0$ and $\mu_0$ being sample thickness, gyromagnetic ratio and vacuum permeability, respectively. It is clear that the diffusion behaviors are different for skyrmion $Q = \pm 1$ [$\mathcal{G}^2 \propto (\pm Q)^2 = 1$] and topologically trivial bubbles $Q = 0$ ($\mathcal{G}^2 = 0$). Namely, $\mathfrak{D}_{dc} = k_\text{B} T/\alpha \mathcal{D}$ is found for $Q = 0$ trivial bubbles, based on which one could determine $\alpha$ and $\mathcal{D}$ by measuring $\mathfrak{D}_{dc}$ *vs. T*. This aspect has been studied in Garnets in which both DMI and uniform topology are irrelevant(*41*). Values of $\mathfrak{D}_{dc}$ are, however, indistinguishable for $Q = \pm 1$ skyrmions which thus cannot be used for quantifying the topology dependent Brownian diffusion, as shown in Fig. 1(D).

**A theoretical approach based on the stochastic Thiele equation.** The quasiparticle nature of skyrmion enables its thermal diffusion to be approximated by considering the motion of its center of effective mass ($m_e$), which results in the stochastic LLG equation to be simplified as the stochastic Thiele equation. Subsequently, the diffusion dynamics of skyrmions and more importantly, their connection with the spin topology will be established (*44, 45*). The stochastic Thiele equation shares a similar form with Langevin equation(*41*), the latter is often utilized to study conventional Brownian particles(*38*), which reads as(*26, 28, 46*):

$$\mathcal{G}\hat{z} \times \boldsymbol{v} + \alpha \boldsymbol{\mathcal{D}} \cdot \boldsymbol{v} + m_e \dot{\boldsymbol{v}} + \alpha \boldsymbol{\Gamma} \times \dot{\boldsymbol{v}} = \boldsymbol{F}_{th} \tag{3}$$

where $\boldsymbol{v}$ is the diffusion velocity, $\boldsymbol{\mathcal{D}}$ is the dissipative tensor(*47*), $\boldsymbol{\Gamma}$ is gyro-damping vector of skyrmions, respectively. The first term on the left side of Eq. (3) implies the topology-dependent diffusion, as a result of the Magnus force that is always orthogonal to the velocity. The second term corresponds to the topology-independent dissipation, the third term represents the acceleration and the fourth term is the skyrmion gyrodamping. $\boldsymbol{F}_{th}$ is the variance of effective force produced by the Gaussian stochastic fluctuations. Assuming the variation of velocity is faster than the change of $\boldsymbol{F}_{th}$, the diffusion velocity can be obtained as follows:

$$\begin{bmatrix} v_x(t) \\ v_y(t) \end{bmatrix} = \frac{1}{\mathcal{G}^2 + \alpha^2 \mathcal{D}^2} \begin{bmatrix} \alpha \mathcal{D} F_{th}^x + \mathcal{G} F_{th}^y \\ \alpha \mathcal{D} F_{th}^y - \mathcal{G} F_{th}^x \end{bmatrix} + \sqrt{v_x^2(0) + v_y^2(0)} e^{Kt} \begin{bmatrix} \cos(Lt) \\ -\sin(Lt) \end{bmatrix} \tag{4}$$

where $v_{x,y}(0)$ are the initial velocities of the skyrmion, $K = -(\alpha \mathcal{D} m_e + \alpha \mathcal{G} \Gamma)/(m_e^2 + \alpha^2 \Gamma^2)$ and $L = (\mathcal{G} m_e - \alpha^2 \mathcal{D} \Gamma)/(m_e^2 + \alpha^2 \Gamma^2)$. The first term on the right side corresponds to the conventional Brownian diffusion while the second term is related to the gyromotion of skyrmions. Depending on the sign of $Q = \pm 1$, skyrmions will gyrotropically move in a clockwise or counterclockwise fashion, with a period $T_p = 2\pi (m_e^2 + \alpha^2 \Gamma^2)/|\mathcal{G} m_e - \alpha^2 \mathcal{D} \Gamma|$. In particular, an average rotation angle $\bar{\theta}_{sr}$ between two neighboring velocity vectors can be used to quantify the gyromotion as follows:

$$\bar{\theta}_{sr} = \langle \arcsin \frac{[\boldsymbol{v} \times (\boldsymbol{v} + d\boldsymbol{v})]_z}{|\boldsymbol{v}||\boldsymbol{v} + d\boldsymbol{v}|} \rangle \approx -\frac{\mathcal{G} m_e - \alpha^2 \mathcal{D} \Gamma}{m_e^2 + \alpha^2 \Gamma^2} \Delta t \tag{5}$$

Since the sign of gyrocoupling vector $\mathcal{G}$ is determined by the sign of $Q$, $\bar{\theta}_{sr} < 0 \, (> 0)$ can be found for $Q = +1 \, (-1)$ skyrmions, given a small damping parameter(*26, 28*). In particular, the amplitude of gyromotion of skyrmions will gradually damp to the equilibrium position for system with a damping parameter $\alpha = 0.02$, as shown in Figs. 1(E) and 1(F). Thus, the gyromotion can be used to reveal the *topology-dependent Brownian gyromotion*. Based on the trajectories shown in Figs. 1(A) and 1(B), a distinct difference of $\bar{\theta}_{sr}$ for $Q = \pm 1$ can be seen in Fig. 1(G), which will be tested experimentally.

**Experimental revelation of skyrmion Brownian gyromotion.** The Brownian motion of a single isolated Néel-type skyrmion is studied by using a polar magneto-optic Kerr effect (MOKE) microscope in an interfacially asymmetric Ta(5nm)/CoFeB(1nm)/TaO$_x$(3nm) multilayer. In our asymmetric multilayer, an interfacial DMI stabilizes uniform Néel-type skyrmions with a left spin chirality(*13*). The film exhibits a relatively weak perpendicular magnetic anisotropy ($H_k = 45$ mT) that is accompanied with stripe to skyrmion morphological transition upon applying magnetic fields(*48, 49*). Note that spin fluctuations and Brownian motion in other skyrmion materials have been previously identified(*29, 49, 50*), the onset of topology-dependent Brownian gyromotion is, however, not discussed. This is the key difference with magnetic bubbles stabilized by the dipolar interaction in Garnets(*51*). To study the free Brownian gyromotion, we have stabilized a single isolated skyrmion of radius $r_{\text{sk}} \sim 1.30$ μm in a disk with a diameter of 130 μm. This enables the complication due to the skyrmion-skyrmion interaction and the asymmetric confining effect to be prohibited. The radius of the experimentally observed skyrmion is much larger than that of the simulated one, as a result of the strong dipole-dipole interaction, which introduces enhanced dissipation as well as skyrmion-defect interaction(*51*). Opposite skyrmions ($Q = \pm 1$) are stabilized by reversing the polarity of perpendicular magnetic fields from positive to negative(*22*).

The trajectory of a single skyrmion ($Q = -1$) at $T = 320.6$ K and $B_z = 2.1$ mT is shown in Fig. 2(A). The random-walk trajectory reflects the complex energy landscape settled by the randomly distributed material defects(*52*), in which the skyrmion constantly picks up kinetic energy from thermal fluctuation that results in escaping from the local energy minima. Shown on the right side of Fig. 2(B) is a MOKE image taken during the diffusion of skyrmion. By summing up the times of the skyrmion appearing at different locations ($N_s$) and divided by the total number of images ($N_t$), it can be seen that the probability ($N_s/N_t$) of finding the skyrmion at certain location is larger than the other locations, as shown in Fig. 2(C). This can be attributed to the pinning effects of the randomly distributed defects.

To understand the dynamical feature of the skyrmion diffusion, we define a *transient velocity* $\boldsymbol{v}(t) = [v_x(t), v_y(t)]$ with $v_x(t) = [X(t + \Delta t) - X(t)]/\Delta t$ and $v_y(t) = [Y(t + \Delta t) - Y(t)]/\Delta t$, with $\Delta t$ being the time interval between two consecutive images limited by the temporal resolution of MOKE microscope, $t^* = N \cdot \Delta t$ where $N$ is an integer. While the measured $v_x(t), v_y(t)$ fluctuate as a function of time, they reside around zero, as shown in Figs. 2(D) and 2(E), respectively. This can be understood as follows, while the force provided by random thermal fluctuation is dynamically varying as a function of time, which, on average is zero, *i.e.*, $\langle h_i(\boldsymbol{x}, t) \rangle = 0$. Meanwhile, it is found that the distribution of the transient velocities and probability density functions (PDFs) can be fitted by a Gaussian distribution, as shown in Figs. 2(F) and 2(G). This is enabled by studying the 1-D PDF which is defined as PDF = $1/\sigma\sqrt{2\pi} \cdot \exp[-(x - \mu)^2/2\sigma^2]$, where $\mu$ and $\sigma$ are the mean and standard deviation of the measured transient velocities, respectively(*36, 38*).

Subsequently, the full width at half maximum (FWHM) of PDFs of $Q = -1$ skyrmion are acquired as $\bar{v}_{x,\text{FWHM}} = (23.3 \pm 0.7)$ μm/s, $\bar{v}_{y,\text{FWHM}} = (24.4 \pm 0.8)$ μm/s, respectively. For $Q = +1$ skyrmion, these values are $\bar{v}_{x,\text{FWHM}} = (24.0 \pm 0.7)$ μm/s, $\bar{v}_{y,\text{FWHM}} = (23.9 \pm$ 0.6) $\mu$m/s, as given in Supplemental Material(*43*). It should emphasize here that the attempt to use the (average) diffusion skyrmion Hall angle $\tan\theta = \bar{v}_{y,\text{FWHM}}/\bar{v}_{x,\text{FWHM}}$ to quantify the topology-dependent skyrmion diffusion are inappropriate, since the topology dependent information will be averaged out in the skyrmion Hall angle ($\bar{v}_{y,\text{FWHM}}/\bar{v}_{x,\text{FWHM}} = \sqrt{\langle v_y^2\rangle/\langle v_x^2\rangle}$).

The temperature dependence of Brownian motion is shown in Fig. 3. It can be seen that when $T = 306.3$ K, the skyrmion is pinned at a fixed location, which is due to the amplitude of the thermal fluctuation is not large enough to excite the skyrmion to escape from the pinning potential. When $T > 306.3$ K and increases further, the skyrmion collects sufficient kinetic energy from the thermal fluctuation and becomes thermally active, which results in the increasing amplitudes of the Brownian motion, as shown in Figs. 3(A) – 3(C). Note that when the temperature is larger than a certain threshold (*i.e.*, $T > 330$ K), the static skyrmion phase is unfavorable and accompanied with creation and annihilation of skyrmions induced by thermal fluctuations. Based on the temperature-dependent skyrmion trajectories, the corresponding MSDs can be calculated, as shown in Fig. 3(D). From the linear regime of MSD *vs.* $t^*$ (we choose $N < 280$)(*53*), $\mathfrak{D}_{dc}$ at various temperatures are shown in Fig. 3(E). It can be seen that the evolution of $\mathfrak{D}_{dc} - T$ are largely overlapping for $Q = \pm 1$ skyrmions, as suggested by Eq. (2). On the other hand, the diffusion coefficient $\mathfrak{D}_{dc} < 4 \times 10^{-11}$ m$^2$s$^{-1}$ is obtained over the whole temperature range, which is much smaller than the value of $\mathfrak{D}_{dc} = \sim 3.0 \times 10^{-8}$ m$^2$s$^{-1}$ obtained from simulations. We note that the experimentally observed skyrmion radius is approximately 100 times larger than the simulated skyrmion radius, and the diffusion coefficient $\mathfrak{D}_{dc}$ given by Eq. (2) thus rapidly decreases with the increase of skyrmion radius since it contributes significantly to the dissipative tensor $\mathcal{D}$.

In particular, a nonlinear $\mathfrak{D}_{dc} - T$ relation is experimentally observed in a narrow temperature range ($306\text{ K} \leq T \leq 330\text{ K}$), which is consistent with our micromagnetic simulation. In our films, the diffusion of the skyrmion is strongly influenced by pinning, especially at lower temperatures (*e.g.*, $T = 300$ K), this breaks the linear correlation predicted by Eq. (2). Considering the combined contribution from the intrinsic free Brownian motion (linear form) and thermally activated depinning (Arrhenius form), the temperature-dependent evolution of diffusion coefficient can be modified as(*29*):

$$\mathfrak{D}_{dc} = CT \exp\left[-\frac{E_p}{k_B(T-T_0)}\right] \qquad (6)$$

where $C = \alpha \mathcal{D} k_B / [\mathcal{G}^2 + (\alpha \mathcal{D})^2]$ and $E_p$ is pinning energy, respectively. Here $T_0 = 305.3$ K, is the activation temperature, above which skyrmion is thermally activated from the local pinning potential and exhibit a Brownian gyromotion. The green curve in Fig. 3(E) is the fitting result, from which we can get $C = 1.29 \times 10^{-12}$ m$^2$/(K·s) and the pinning energy $E_p \approx$ 8.2 meV = 94.7 K.

The experimental demonstration of topology-dependent Brownian gyromotion is presented in Fig. 4. From the time-dependent trajectories at $T = 328.5$ K, clockwise and counterclockwise rotations of a single skyrmion can be seen for $Q = +1$ and $Q = -1$ skyrmions in Figs. 4(A) and 4(B), respectively. For skyrmions with opposite topological charges, the evolution of rotation angle as a function of temperature is systematically calculated and shown in Fig. 4(C), in which a clear difference in the whole temperature range is evident. Note that the observed (average) skyrmion rotation angle $\bar{\theta}_{sr}$ is relatively small, which can be attributed to the strong pinning effect that gives rise to hopping between different pinning sites and results in a deviation from the free gyromotion. Our experimental results are consistent with the analytical calculations from the stochastic Thiele equation, suggesting the Brownian gyromotion of a single skyrmion is being captured.

In summary, we have numerically and experimentally studied the Brownian diffusion of a single isolated skyrmion driven by the random thermal fluctuations. The characteristics of skyrmion, including random trajectory and Gaussian distribution of probability density function, are consistent with conventional Brownian particles. Moreover, by studying skyrmions with opposite spin topology ($Q = \pm 1$), we have observed a topology-dependent gyromotion. Our results thus revealed the intriguing topological aspects of nonequilibrium thermodynamics of skyrmions, that can be potentially extended for studying other topological spin textures, antiskyrmions(*54*) or merons(*55*), for examples. It is also expected that our results could provide useful insights for developing thermally stable skyrmion-based spintronic devices and potentially enable novel applications of skyrmions in reservoir/probability computing schemes(*34, 35*).

**Acknowledgements**. Work carried out at Tsinghua University was supported by the Basic Science Center Project of NSFC (Grant No. 51788104), National Key R&D Program of China (Grant Nos. 2017YFA0206200 and 2016YFA0302300), the National Natural Science Foundation of China (Grant No. 11774194, 51831005, 1181131008, 1181101082), Beijing Natural Science Foundation (Grant No. Z190009), Tsinghua University Initiative Scientific Research Program and the Beijing Advanced Innovation Center for Future Chip (ICFC). X.Z.

was supported by the Guangdong Basic and Applied Basic Research Foundation (Grant No. 2019A1515110713), and the Presidential Postdoctoral Fellowship of The Chinese University of Hong Kong, Shenzhen (CUHKSZ). Y.Z. acknowledges the support by the President's Fund of CUHKSZ, Longgang Key Laboratory of Applied Spintronics, National Natural Science Foundation of China (Grant Nos. 11974298 and 61961136006), Shenzhen Fundamental Research Fund (Grant No. JCYJ20170410171958839), and Shenzhen Peacock Group Plan (Grant No. KQTD20180413181702403).

**Figures**

FIG. 1 (color online). Micromagnetic simulation of skyrmion Brownian gyromotion. Simulated trajectories of a single isolated skyrmion ($Q = +1$) and ($Q = -1$) with its radius $r_{sk} <$ 100 nm at 240 K, as shown in Figs. (A) – (B). The total simulation time is 18 ns with a step of 20 ps. (C) MSDs of $Q = \pm 1$ skyrmions. (D) Diffusion coefficients as a function of temperature ($\mathfrak{D}_{dc}$ vs. $T$) for $Q = \pm 1$ skyrmions. (E) and (F) Gyrotropic diffusion of skyrmions with $Q = \pm 1$ for a damping coefficient ($\alpha = 0.02$). (G) Evolution of the average skyrmion rotation angle $\bar{\theta}_{sr}$ as a function of temperature for $Q = \pm 1$ skyrmions.

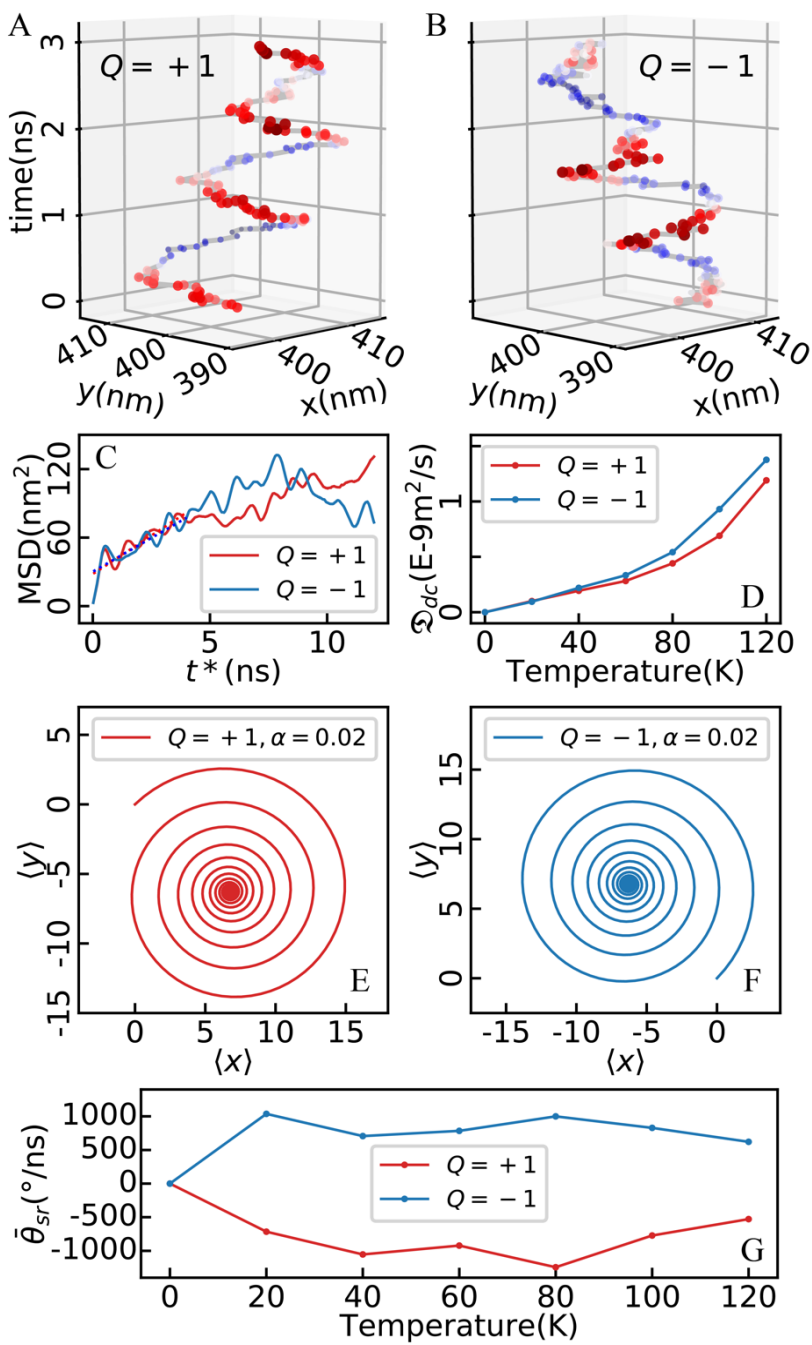

FIG. 2 (color online). Brownian motion of a single isolated skyrmion ($Q = -1$) in a Ta/CoFeB/TaO$_x$ disk at $T = 320.6$ K and $B_z = +2.1$ mT. (A) Trajectory of the skyrmion captured from the video tracking results with 28 frames per second over 957s. MOKE image of the device is shown in (B). Dark contrast corresponds to the magnetization pointing along the $+z$ direction. Left-handed coordinate originates from data-readout process when using ImageJ. (C) Spatially varying skyrmion appearing probability. The boundary of device is marked as the black circle. (D) and (E) Measured $v_x$, $v_y$ as a function of $t$. (F) and (G) Probability density functions (PDFs) of $v_x$ and $v_y$, respectively. Values of FWHM were obtained from Gaussian fittings.

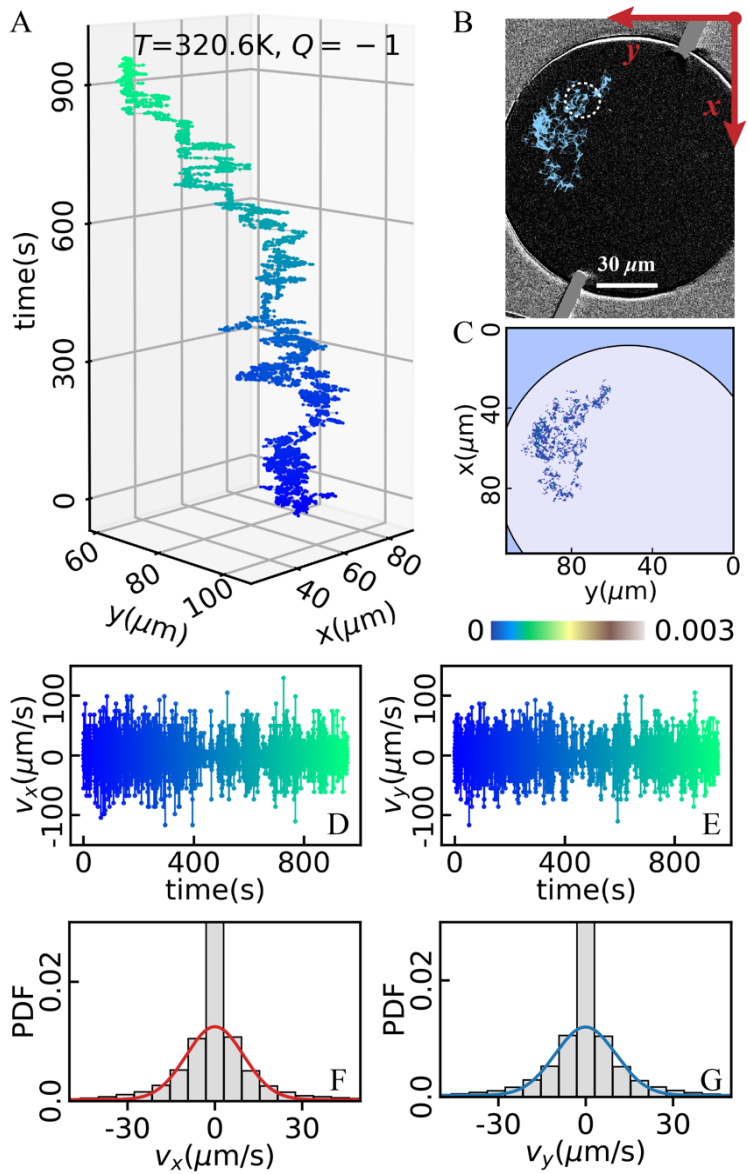

FIG. 3 (color online). Evolution of trajectories of a single isolated ($Q = -1$) skyrmion at various temperatures, (A) $T = 306.3$ K, (B) $T = 314.9$ K, (C) $T = 326.5$ K. (D) The calculated MSDs of $Q = -1$ skyrmion measured at different temperatures. (E) The dependence of $\mathfrak{D}_{dc}$ vs. $T$ for $Q = \pm 1$ skyrmions. Values of $\mathfrak{D}_{dc}$ at different temperatures were calculated based on the MSD data by using Eq. (1), which can be fitted by using Eq. (6) (green solid line).

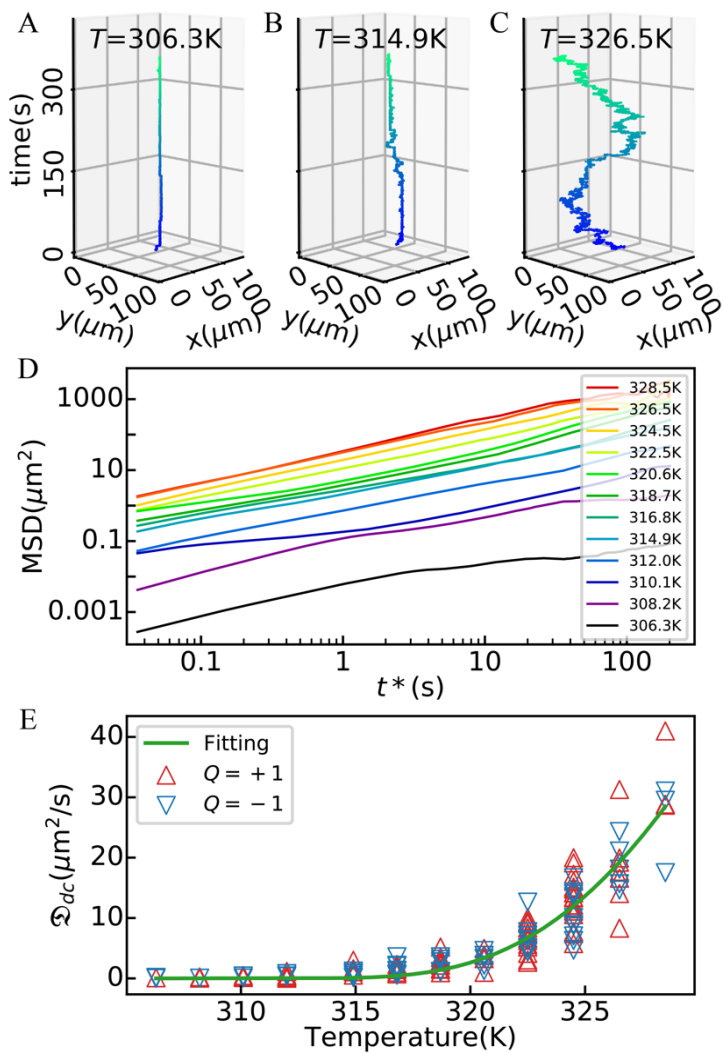

FIG. 4 (color online). Gyrodiffusion of skyrmions at $T = 328.5$ K were shown for (A) $Q = +1$, and (B) $Q = -1$, respectively. (C) The evolution of average skyrmion rotation angle $\bar{\theta}_{sr}$ as a function of temperatures, in which opposite values of $\bar{\theta}_{sr}$ for ($Q = \pm 1$) can be found over the whole temperature range. Each data point is obtained by averaging approximately 20000 frames.

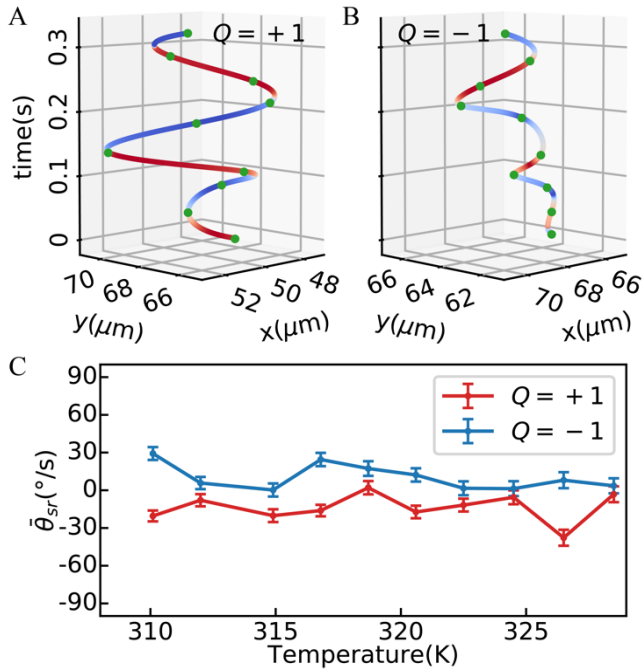